\documentclass[12pt]{article}
\newcommand{\BEQ}{\begin{equation}}
\newcommand{\EEQ}{\end{equation}}
\newcommand{\sphi}[3]{\frac{\sigma(#1#2#3)}{\sigma(#3)\sigma(#1)}}
\newtheorem{pr}{Theorem}

\makeatletter
\@addtoreset{equation}{section} \makeatother

\usepackage{amsfonts}
\usepackage{amssymb}
\begin{document}

\begin{titlepage}

\hfill ITEP-TH-04/01
\vskip 1.5cm
\centerline{\Large \bf
  Elliptic Gaudin system with spin }
\vskip 1.0cm \centerline{D. Talalaev \footnote{E-mail:
talalaev@gate.itep.ru} } \centerline{Institute for Theoretical and
Experimental Physics \footnote{ITEP, Moscow, Russia.}} \vskip
3.0cm {\bf Abstract.} The elliptic Gaudin model was obtained
\cite{NN,ER1} in the framework of the Hitchin system on an
elliptic curve with punctures. In the present paper the
algebraic-geometrical structure of the system with two fixed
points is clarified. We identify this system with poles dynamics
of the finite gap solutions of Davey-Stewartson equation. The
solutions of this system in terms of theta-functions and the
action-angle variables are constructed. We also discuss the
geometry of its degenerations.


\end{titlepage}

\section*{Introduction}
The integrable structure of the Calogero-type systems is of
special interest concerning such diverse problems as hierarchies
of partial differential equations \cite{K1}, Seiberg-Witten
solutions of N=2 SUSY \cite{GK,DW}, quasi-Hopf deformations of
quantum groups and many others. There are many different
approaches to describe the Hamiltonian structure of such systems.
One of the aims of this paper is to combine the Hitchin
description and the language of the universal algebraic-geometric
symplectic form. The method of the special inverse spectral
problem was successfully applied in the matrix KP case in
\cite{KBBT}. This method deals with the auxiliary linear problem
coming from the Lax representation of the equation. The
Baker-Akhiezer function (FBA) turns to be a function on the
spectral algebraic curve, and its specific analytic properties are
very restrictive and provide the reconstruction of the potential
and the poles of FBA in terms of theta-functions on the Jacobian
of the spectral curve. The same procedure was applied to the spin
Ruijsenaars-Schneider system in \cite{KZ}. But for a wide class of
integrable systems of the Hitchin type the explicit solutions
remain still unknown. In the first section we revisit the
construction of the Hitchin system on the degenerate curve and
especially the Gaudin elliptic two-point N-particle system with
spin. In the second section we investigate the finite gap
potentials of the matrix Davey-Stewartson equation. The system of
non-linear equations of its poles dynamics is deduced and the
algebraic-geometrical correspondence between the phase space of
this system and the space of spectral data is constructed. In the
third part the FBA is reconstructed in terms of
$\theta$-functions, the explicit expressions for the potential and
for the solution of the poles dynamics are given. The universal
symplectic structure is constructed in the fourth section. We also
demonstrate that the system of poles dynamics is hamiltonian and
identify it with the Gaudin elliptic system with spin. In the last
section we discuss the rational and special modular degenerations
of the system.

\vskip 1.0cm

\section{Gaudin system}
The Hitchin system was introduced in \cite{Hit} as an integrable
system on the cotangent bundle of the moduli space of stable
holomorphic bundles on the algebraic curve $\Sigma$. The phase
space was obtained by the Hamiltonian reduction from the space of
pairs $d_A^{''},\Phi$, where $d_A^{''}$ is the operator defining
the holomorphic structure on the bundle $V$ and $\Phi$ is the
endomorphism of this bundle, more precisely $\Phi \in
\Omega^{0,1}(\Sigma,End(V))$. The reduction by the action of the
loop group $GL_N(z)$ was done, i.e. the group of $GL_N$-valued
functions on $\Sigma$. The invariant symplectic structure on the
base space writes \BEQ \omega=\int_{\Sigma}Tr \delta \Phi \wedge
\delta d_A^{''}. \label{sym} \EEQ The zero level of the moment map
is described by the condition $d_A^{''} \Phi=0$ which means that
$\Phi$ is holomorphic with respect to the induced holomorphic
structure on the bundle $End(V)$. It turns out that the system of
quantities $Tr\Phi^k$, treated as vector functions on the phase
space, Poisson-commute and their number is exactly half the
dimension of the phase space. This system was stated in a formal
way and its explicit formulation was obtained for special cases in
\cite{GP,GAW}.

The generalized construction of the
Hitchin systems on degenerate curves \cite{NN} provides a more
explicit description. One considers the moduli space of semi-stable
bundles $V$ on an algebraic curve with singularities.
The Higgs field in this case is a meromorphic section
of the bundle $End(V)$. The holomorphic
bundle on the singular curve is the holomorphic bundle on its
normalization and the set of additional data characterizing
the bundle over singularities.

Let us consider the elliptic curve with two simple fixed points $z_1,z_2$.
The Higgs field is a meromorphic section with two poles on the curve.
The moduli space $\mathcal M$ of semi-stable bundles on the elliptic curve
with such type of singularity decomposes into the moduli space of
semi-stable bundle $V$ on the normalized
curve and the finite dimensional space of
additional data at the fixed points: $g_i\in End(V|_{z_i})$. The cotangent
bundle ${\mathcal T}^*\mathcal M$ has additional coordinates
$p_i \in {\mathcal T}^*End(V|_{z_i})$ where $i=1,2$. The modified symplectic structure is
\BEQ
\omega=\int \delta \Phi \wedge \delta d_A^{''} +Tr (\delta (g_1^{-1}p_1) \wedge
\delta g_1 + \delta (g_2^{-1}p_2)\wedge \delta g_2)
\label{sym1}
\EEQ
and the zero level of the modified moment map is
\BEQ
\bar \partial \Phi +[A,\Phi]+p_1 \delta^2(z_1)+p_2 \delta^2(z_2)=0.
\label{mom}
\EEQ
We restrict ourselves to the subspace of semi-stable holomorphic bundles
of rank $N$
which decompose into the sum of line bundles.
In the basis associated
to the decomposition we can make the matrix $A$ diagonal
with elements $x_1,\ldots,x_n$. The equation
(\ref{mom}) becomes
\BEQ
\bar \partial \Phi_{ij}+(x_i-x_j)\Phi_{ij}+(p_1)_{ij}\delta^2(z_1)+
(p_2)_{ij}\delta^2(z_2)=0.
\EEQ
It could be solved as in \cite{NN} in the following way
\BEQ
\Phi_{ij}=\frac {\exp((x_i-x_j)\frac{z-\bar z}{\tau-\bar \tau})}{2 \pi i}
((p_1)_{ij}\frac {\sigma(x_i-x_j+z-z_1)}{\sigma(x_i-x_j) \sigma(z-z_1)}+
(p_2)_{ij}\frac {\sigma(x_i-x_j+z-z_2)}{\sigma(x_i-x_j) \sigma(z-z_2)});
\EEQ
\BEQ
\Phi_{ii}=\omega_i+(p_1)_{ii} \zeta(z-z_1) +(p_2)_{ii} \zeta(z-z_2);
\EEQ
where $\tau$ is the modular parameter of the elliptic curve,
$\sigma$ is the Weierstra\symbol{25} $\sigma$-function and $\omega_i$ are
additional free parameters.
Now $res_{z=z_1}4\pi^2 Tr \Phi^2$ is equal to
\BEQ
H_1=\sum_i \omega_i (p_1)_{ii} +\sum_i (p_1)_{ii} (p_2)_{ii} \zeta(z_1-z_2)
+\sum_{i \ne j} (p_1)_{ij} (p_2)_{ji}
\frac{\sigma(x_i-x_j+z_1-z_2)}{\sigma(x_i-x_j) \sigma(z_1-z_2)},
\label{ham}
\EEQ
and the residue at the point $z=z_2$ gives a similar
expression for the Hamiltonian $H_2$ which differs from
$H_1$ by the interchanging of $z_1$ and $z_2$.
These functions are called the Hamiltonians of the Gaudin
elliptic system with spin.
We are interested in the complete integrability of this system.
On the moduli space of
holomorphic bundles on the regular curve in  \cite{Hit}
the number of coordinates on the set of quantities $Tr \Phi^k$, i. e.
the number of Beltrami differentials, was proved to be exactly half
the dimension of $\mathcal T \mathcal M$. Our case differs from the classical one and we calculate
the dimension directly.  The
coefficients of the principal part of
$Tr \Phi^k$ at the vicinity of the fixed points are the integrals.
The mentioned
quantities are elliptic functions, so the sum of their residues is equal
to zero.
The number of integrals is equal to $N(N+1)$. The number of
coordinates on the
phase space $x_i,\omega_i(p_1)_{ij},(p_2)_{ij},(g_1)_{ij},(g_2)_{ij}$ is equal
to $2N+4N^2$. The symplectic structure (\ref{sym1}) is invariant under the
action of the group $GL_N\times GL_N$ in the following way: the element
$(g^0_1,g^0_2)$ acts as
$$
g_1\mapsto g_1 g_1^0,\qquad g_2 \mapsto g_2 g_2^0.
$$
The moment map of this action is $\mu=((g_1)^{-1}p_1 g_1,
(g_2)^{-1}p_2 g_2)$ and we can fix it to have a diagonal form and
this kills $2N(N-1)$ degrees of freedom. The stabilizer of
this choice of moment map consists of the diagonal subgroup
$Diag \times Diag \in G \times G.$ It means that
the choice of the representative of the stabilizer's orbit
in the moment map subspace kills $2N$ additional degrees of freedom.
The dimension of the reduced phase space is equal to
$2N+4N^2-2N(N-1)-2N=2N(N+1)$ so that the number of integrals is
exactly half the number of variables.

\section{Davey-Stewartson equation}
As was shown in \cite{Mal}, the Davey-Stewartson equation
\BEQ
iu_t + u_{xx} - u_{yy} +u \phi = 0 \qquad \phi_{xx} +
\phi_{yy} =
2 ( {|u|}_{xx}^2 - {|u|}_{yy}^2 )
\label{DS2}
\EEQ
has a Lax representation $\dot L=[L,M]$ with
$$ L= \left( \matrix{ \partial_z & 0 \cr 0 & \partial_{\bar z}
 \cr } \right) -
\frac 1 2  \left( \matrix{ 0 & u \cr v & 0 \cr } \right) \qquad
z=x+iy .$$ Let us consider the $l$-dimensional generalization of
the auxiliary linear problem related to this operator \BEQ \left(
\partial_t + \left( \matrix{ d_1 & 0 \cr 0 & d_2 \cr } \right)
\partial_x + \left( \matrix{ 0 & A_1 \cr A_2 & 0 \cr } \right)
\right) \Psi =D_A \Psi=0, \label{LP} \EEQ where $A_1,A_2$ are
$l\times l$-matrices, $d_1=E,d_2=-E$, and $E$ is the identity
$l\times l$-matrix, and $\Psi$ is a $2l$-dimensional vector
function. Looking for the quasi-periodic solutions of the matrix
Davey-Stewartson equation we put the potential in the special
form: \BEQ A_1=\sum_{j=1}^n a_1^j {b_1^j}^+ \frac
{\sigma(x-x_j+\eta)}{\sigma(\eta) \sigma(x-x_j)}, \qquad
A_2=-\sum_{j=1}^n a_2^j {b_2^j}^+ \frac {\sigma(x-x_j-\eta)}
{\sigma(\eta) \sigma(x-x_j)}, \label{pot} \EEQ where $a_k^j$ and
$b_k^j$ are $l$-vectors, and $\sigma$ is a Weierstra\symbol{25}
function. As usual we look for the quasi-periodic solutions of the
linear problem. The convenient basis of such functions with simple
poles on $x$ is provided by $$\Phi_i(z,x) =\frac {\sigma(z-z_i+x)}
{\sigma(x) \sigma(z-z_i) } \exp(-\zeta(z) x),\qquad
\mbox{where}\qquad z_2-z_1=\eta.$$ We call $\Psi_1$ the first $l$
components of the function $\Psi$, and $\Psi_2$ the last $l$
components. The blocks of the potential have a monodromy such that
the ratio of the monodromies of $\Psi_1$ and $\Psi_2$  must be
equal to the monodromy of $A_1$. We look for the eigenfunction in
the form: \BEQ \label{anz} \Psi_i=\sum_{j=1}^n S_i^j
\Phi_i(z,x-x_j) \exp( k y_+ +\frac 1k y_-), \EEQ where $y_{\pm}
=\frac 12 (t \pm x)$. We also look for the solution of the dual
equation \BEQ \label{LP+} \Psi^+ D_A =0 \EEQ in the form \BEQ
\label{anz+} \Psi_i^+=\sum_{j=1}^n {S_i^j}^+ \Phi_i(z,x_j-x)
\exp(-k y_+ -\frac 1k y_-). \EEQ
\begin{pr} The equation (\ref{LP}) has a solution of the form (\ref{anz})
and the equation (\ref{LP+}) has a solution of the form (\ref{anz+}) if and
only if the poles dynamics is described by the system:
\BEQ
\dot a_1^j+\lambda_1^j a_1^j + \sum_{l \ne j} a_1^l ({b_1^l}^+ a_2^j)
\sphi{x_j-x_l}{+}{\eta}=0;
\label{El1}
\EEQ
\BEQ
\dot a_2^j+\lambda_2^j a_2^j - \sum_{l \ne j} a_2^l ({b_2^l}^+ a_1^j)
\sphi{x_j-x_l}{-}{\eta}=0;
\label{El2}
\EEQ
\BEQ
\dot b_1^j-\lambda_2^j b_1^j + \sum_{l \ne j} b_1^l ({b_2^j}^+ a_1^l)
\sphi{x_j -x_l}{+}{\eta}=0;
\label{El3}
\EEQ
\BEQ
\dot b_2^j-\lambda_1^j b_2^j - \sum_{l \ne j} b_2^l ({b_1^j}^+ a_2^l)
\sphi{x_j-x_l}{-}{\eta}=0;
\label{El4}
\EEQ
\BEQ
\ddot x_j = \sum_{l \ne j}
\left( ({b_1^l}^+ a_2^j)({b_2^j}^+ a_1^l)\sphi{x_j-x_l}{+}{\eta} -
 ({b_1^j}^+ a_2^l)({b_2^l}^+ a_1^j)\sphi{x_j-x_l}{-}{\eta} \right);
\label{El5}
\EEQ
$$
\dot \lambda_1^j- \dot \lambda_2^j =\sum_{l \ne j}
 ({b_2^j}^+ a_1^l)({b_1^l}^+ a_2^j)\sphi{x_j-x_l}{+}{\eta}
(\zeta(x_j-x_l)-\zeta(x_j-x_l+\eta)+2\zeta(\eta)) +
$$
\BEQ
\sum_{l \ne j} ({b_1^j}^+ a_2^l)({b_2^l}^+ a_1^j)
\sphi{x_j-x_l}{-}{\eta}(\zeta(x_j-x_l)-\zeta(x_j-x_l-\eta)-2\zeta(\eta)).
\label{El6}
\EEQ

\end{pr}
{\bf Proof}
As was done in \cite{KBBT,KZ,Tal}, introducing the expression (\ref{anz})
into equation (\ref{LP}) and eliminating the essential singularity,
we calculate the coefficients of the expansion in $x$ at the
vicinity of the poles $x=x_j$. The lowest term is of order $-2$
$$
S_1^j (\dot { x_j } - 1) + a_1^j { b_1^j }^+ S_2^j = 0;
$$
$$
S_2^j (\dot { x_j } + 1) + a_2^j { b_2^j }^+ S_1^j = 0.
$$
The similar condition from the dual equation gives
$$
{ S_1^j }^+ (\dot { x_j } - 1) +
{ S_2^j }^+a_2^j { b_2^j }^+ = 0;
$$
$$
{ S_2^j }^+ (\dot { x_j } + 1) +
{ S_1^j }^+ a_1^j { b_1^j }^+ = 0.
$$
We can fix the gauge of variables $a_k^j,b_k^j$ such that
$$ 1 - \dot { x_j } = ({b_1^j}^+ a_2^j ), \qquad
-1-\dot { x_j } = ( {b_2^j}^+ a_1^j ),$$
and so
\BEQ
 S_k^j = c^ja_k^j, \qquad { S_k^j }^+ ={ c^j }^+
 b_{ \bar k }^j \qquad \bar k =3-k,
\label{cs}
\EEQ
where $c^j$ are scalar functions.
The terms of order $-1$ at $x=x_j$ implying (\ref{cs})
can be represented as (\ref{El1}),(\ref{El2}) where $\lambda_k^j$
is defined from the following equations on $c^i$
\BEQ
(\partial_t+kI+L_1)C=0;
\label{L1}
\EEQ
\BEQ
(\partial_t+\frac 1 k I +L_2)C=0,
\label{L2}
\EEQ
where
$$ {(L_1)}_{ij} = \delta_{ij}(({b_1^i}^+a_2^i)
(\zeta(z-z_2)-\zeta(z)-\zeta(\eta))-\lambda_1^i)
+(1-\delta_{ij})({b_1^i}^+ a_2^j) \Phi_2(x_i-x_j,z);$$
$$ {(L_2)}_{ij} = \delta_{ij}(({b_2^i}^+a_1^i)
(\zeta(z-z_1)-\zeta(z)+\zeta(\eta))-\lambda_2^i)
+(1-\delta_{ij})({b_2^i}^+ a_1^j) \Phi_1(x_i-x_j,z).$$
Concerning the dual equation, we obtain similar equations
for the variables ${c^i}^+$
and the equations (\ref{El3}),
(\ref{El4}).
Thus for the Lax equation we have
$$\partial_t(L_1-L_2)=[L_1,L_2],$$
which is equivalent to the equations (\ref{El5}),(\ref{El6}).
$\blacksquare$

The standard trick in the method of algebraic-geometrical
direct and inverse problem is that the spectral curve $\Gamma$,
defined by the characteristic equation $det(k-\frac 1 k +L_1-L_2)=0$,
has specific properties and that
$\Psi$-function defines a holomorphic line bundle on the curve,
and its essential singularity depending on time is the deformation
of the fixed line bundle. The analysis is rather traditional and we
venture to outline main propositions on the analytic properties of
eigenfunction and properties of the spectral curve.
We use the notation $L(z)=L_1(z)-L_2(z)$ and $\hat{k}=k-\frac 1 k.$
First, noting that the matrix $res_{z=z_i}L(z)$ differs from the
unity matrix by a matrix of rank $l$
we find the decomposition at the points $z_1,z_2$
\BEQ
R(\hat k,z)=\prod_{i=1}^N (\hat k + \nu^{1,2}_i z^{-1}+ h^{1,2}_i(z))
\label{razl}
\EEQ
where $h^{1,2}_i(z)$ are regular and $\nu^{1,2}_i=1$ if $i>l$.
Using that $\Gamma$ is an $N$-sheeted covering of the base curve $z$ and the
definition of the branch points of the curve, we count the number of branch
points; it is equal to $4Nl-2l(l+1)$, and by Riemann-Hurwitz theorem
the genus of $\Gamma$ is $g=2Nl-l(l+1)+1.$ In virtue of the Lax
representation the parameters of the spectral curve are integrals of motion.
If, in addition, we fix $c(t)|_{t=0}=const$ it can be deduced that the poles
of vector $c$ do not depend on time. In general position the number
of such poles is half the number of branch points, and it is equal to
$2Nl-l(l+1)=g-1.$
The $\Psi$ function has essential singularities at the points over
$z=z_1,z=z_2$ and additional $2l$ poles. Summarizing the analytic properties
of the eigenfunction we obtain
\begin{pr}
The function $\Psi$ is meromorphic on the curve $\Gamma$
of genus $g$ outside $2l$ points $P_i$,
and has $g+2l-1$ poles
$\gamma_1,...,\gamma_{g+2l-1}$ which do not
depend on $t,x$. At the points
$P_i,i=1,...,l$ over $z=z_1$ it has the following decomposition
$$  \Psi_\alpha (x,t,P)=(\chi_0^{\alpha  i} +
\sum_{s=1}^\infty \chi_s^{\alpha  i}  (x,t) z^s)
\exp(\lambda_i(x+t)) \Psi_1(0,0,P)$$
and when $i=l+1,...,2l$ at the points over $z=z_2$ we have
$$  \Psi_\alpha (x,t,P)=(\chi_0^{\alpha  i} +
\sum_{s=1}^\infty \chi_s^{\alpha  i}  (x,t) z^s)
\exp(\lambda_i(x-t)) \Psi_1(0,0,P)$$
where $\lambda_i(z)=(1-\nu_i)z^{-1} - h_i(z)$, $h_i$ are regular
functions in the vicinity of $z=0$, $\nu_i$ are eigenvalues of the matrix $L$ at
the points over $z=z_1,z=z_2$ respectively and $\chi_0^{\alpha  i}$ are
constants.
\end{pr}
The similar proposition on the analytic properties of the
solution of dual equation is true
\begin{pr}
The solution of the dual equation $\Psi^+$ is a meromorphic function
on the same curve $\Gamma$ of genus $g$ outside the points $P_i,i=1,...,2l$
It has $g+2l-1$ poles $\gamma_1^+,...,\gamma_{g+2l-1}^+$ which do not depend
on $t,x$. At the points $P_i,i=1,...,l$ over $z=z_1$
it has the following decomposition
$$\Psi_\alpha ^+ (x,t,P)=(\chi_0^{+,\alpha  i} +
\sum_{s=1}^\infty \chi_s^{+,\alpha  i}  (x,t) z^s)
\exp(-\lambda_i(x+t)) \Psi_1^+(0,0,P),$$
and at the points $P_i$ over $z=z_2$, $i=l+1,...,2l$ one has
$$  \Psi_\alpha ^+ (x,t,P)=(\chi_0^{+,\alpha  i} +
\sum_{s=1}^\infty \chi_s^{+,\alpha  i}  (x,t) z^s)
\exp(-\lambda_i(x-t)) \Psi_1^+(0,0,P),$$
where $\chi_0^{+,\alpha  i}$ are constants.
\end{pr}

\section{Reconstruction formulas}
In this section we demonstrate that the
analytic properties that we found in the
previous section are sufficiently restrictive and allow us to reconstruct
the eigenfunction in terms of the characteristics of the curve,
such as abelian differentials, $\theta$-functions and Abel transform.
The standard proposition about the uniqueness of the function
with the following properties
\begin{enumerate}
\item $\Psi_\alpha (x,t,P)$  has $g+2l-1$ poles $\gamma_1,...,\gamma_{g+2l-1}$
on the curve $\Gamma$ of genus $g$
\item it has essential singularities at the points $P_j,j=1,..,l$
of the form
$$ \Psi_\alpha (x,t,P)= e^{w_j^{-1}(x+t)} (\delta_{\alpha  j} +
\sum_s F_{\alpha  j}^s (x,t)w_j^s)$$
and at the points $P_j,j=l+1,...,2l$ of the form
$$ \Psi_\alpha (x,t,P)= e^{w_j^{-1}(x-t)} (\delta_{\alpha  j} +
\sum_s F_{\alpha  j}^s (x,t) w_j^s)$$
where $w_j(P)$ is the local parameter at $P_j$
\end{enumerate}
arises from the Riemann-Roch theorem. And its
existence follows from the construction: let $d\Omega^1,d\Omega^2$
be unique meromorphic differentials, holomorphic
outside the points $P_j$ such that:
$$d\Omega^1=d(w_j^{-1}+O(w_j)) \mbox\ {at\ the\ points}\ P_j$$
$$d\Omega^2=\cases{
d(w_j^{-1}+O(w_j)) &at the points $P_j$ with $j=1,...,l$\cr
d(-w_j^{-1}+O(w_j)) &at the points $P_j$ with $j=l+1,...,2l$\cr}
$$
normalized by the condition $\int_{a_j} d\Omega^k=0$.
We set $U_j^{(k)}=\frac {1} {2\pi i}\int_{b_j} d\Omega^k$; then,
up to a normalization constant
\BEQ
\Psi_\alpha (x,t,P)=\frac
{\theta(A(P)+U^{(1)}x+U^{(2)}t+Z_\alpha )\theta(Z_0)
\prod_{j\ne \alpha } \theta(A(P)+S_j)}
{\theta(A(P)+Z_\alpha )\theta(U^{(1)}x+U^{(2)}t+Z_0)
\prod_{i=1}^{2l} \theta(A(P)+R_i)}
e^{x\Omega^{(1)}(P)+t\Omega^{(2)}(P)}
\label{eq:410}
\EEQ
where
$A(P)$ is the Abel transform, $K$ is the vector
of Riemann constants,
$$R_i = -K -\sum_{s=1}^{g-1} A(\gamma_s)-A(\gamma_{g-1+i}),\qquad
S_i = -K-\sum_{s=1}^{g-1} A(\gamma_s) - A(P_i),$$
$$Z_0=-K-\sum_{i=1}^{g+2l-1}A(\gamma_i)+\sum_{j=1}^{2l}A(P_j),
\qquad Z_{\alpha }=Z_0-A(P_{\alpha }),$$
$\Omega^{(k)}=\int_{P_0}^P d\Omega^k$.
The dual eigenfunction is constructed analogously
$$\Psi_\alpha ^+ (x,t,P)=\frac
{\theta(A(P)-U^{(1)}x-U^{(2)}t+Z_\alpha ^+)\theta(Z_0^+)
\prod_{j\ne \alpha } \theta(A(P)+S_j^+)}
{\theta(A(P)+Z_\alpha ^+)\theta(U^{(1)}x+U^{(2)}t-Z_0^+)
\prod_{i=1}^{2l} \theta(A(P)+R_i^+)}
e^{-x\Omega^{(1)}(P)-t\Omega^{(2)}(P)}$$
where
$$S_j^+=-K-A(P_j)-\sum_{s=1}^{g-1}A(\gamma_s^+),\qquad
R_i^+=-K-A(\gamma_{g-1+i}^+),$$
$$K_0=\sum_{i=1}^{g+2l-1}(A(\gamma_i)+A(\gamma_i^+))-
2\sum_{j=1}^{2l}A(P_j),\qquad Z_0^+=Z_0 -2K-K_0,
\qquad Z_\alpha ^+=Z_0^+-A(P_\alpha ).$$
These functions satisfy the auxiliary linear problems (\ref{LP}),
(\ref{LP+}). The finite gap potential of the Dirac operator,
i.e. the solution of the matrix
Davey-Stewartson equation is given by $A=\sigma F^1 \sigma -F^1$,
where $\sigma$ is the diagonal matrix with
diagonal elements $\underbrace{1,\ldots,1}_l,\underbrace{-1,\ldots,-1}_l$
and $F_1$ is the matrix coefficient of the decomposition
near the singularities of the FBA.
Therefore the solution of the system of poles dynamics
(\ref{El1}),...,(\ref{El6}) is given  by the formulas:
\begin{enumerate}
\item the poles of the eigenfunction can be found as
the solutions of the equation
$$ \theta(U^{(1)}x_i(t)+U^{(2)}t+Z_0)=0,$$
\item the spin variables are given by:
$$ \tilde a_{i,\alpha }(t)=Q_i^{-1}(t) \frac
{\theta(U^{(1)}x_i(t)+U^{(2)}t+Z_\alpha )}
{\theta(Z_\alpha ) \theta(S_\alpha)},$$
$$ \tilde b_i^\alpha (t)=Q_i^{-1}(t) \frac
{\theta(U^{(1)}x_i(t)+U^{(2)}t-Z_\alpha ^+)}
{\theta(Z_\alpha ^+) \theta(S_\alpha^+)},$$
\item the normalizing factor is:
$$ Q_i^2(t)=\frac 1 2 \sum_{\alpha =1}^{2l} \frac
{\theta(U^{(1)}x_i(t)+U^{(2)}t-Z_\alpha )  \theta(U^{(1)}x_i(t)+U^{(2)}t-Z_\alpha ^+) }
{\theta(Z_\alpha ) \theta(Z_\alpha ^+) \theta(S_\alpha) \theta(S_\alpha^+)}. $$
\end{enumerate}
In the formulas above we used the notation:
$$ a_{i,\alpha }^1=2 \tilde a_{i,\alpha }, a_{i,\alpha }^2=2 \tilde a_{i,\alpha +l}$$
$$ {b_i^\alpha }^1=\tilde b_i^{\alpha +l}, {b_i^\alpha }^2=\tilde b_i^\alpha. $$

\section{Universal symplectic structure}
The systems obtained as poles dynamics in the finite gap theory were often
introduced from physical motivations. Such an explanation leaves the
structural questions concerning dynamical systems apart. By structure we mean
the algebraic symmetries in the broad sense. The importance of
the hamiltonian structure cannot be overestimated in the
fundamental problem of quantization, likewise the ``action-angle'' variables
have a crucial meaning in the Seiberg-Witten theory.
Heuristically most of
the systems involved are hamiltonian and moreover integrable in the sense
of Liouville.
In \cite{KP,KP1} was proposed the universal hamiltonian description
of such systems in terms of the flows on the space of quasi-periodic
operators generated by the Lax equation. First, this approach was
applied to the description of the hamiltonian structure of the
nonlinear equations of the soliton theory, but in \cite{K2} the
efficiency of this method when applied to the finite-dimensional
case was confirmed.
The essential point is the
algebraic-geometric correspondence which maps the coordinates of the
phase space of the system of poles dynamics to the space of spectral
data, namely to the set of modular coordinates, the divisor of poles of the
Baker-Akhiezer function and the type of singularities, which are given
by a pair of Abelian differentials on the curve, in our case by
the differentials $dk,dz$.

The coordinates on the phase space of the
system (\ref{El1}),...,(\ref{El6}) given by
$x_i,\lambda_i,a_1^j,a_2^j,b_1^j,b_2^j$ are invariant
under the following change of variables
$$ a_1^i \mapsto W_1^{-1}a_1^i,\qquad
{b_2^i}^+ \mapsto {b_2^i}^+  W_1,$$
$$ a_2^i \mapsto W_2^{-1} a_2^i,\qquad
{b_1^i}^+ \mapsto {b_1^i}^+  W_2$$
where $W_k,k=1,2$ are constant matrices. We also have the symmetry of the
system
$$ a^i \mapsto \mu_i a^i;\qquad b^i \mapsto {\mu_i}^{-1} b^i$$
where $\mu_i$ are constant scalars.
We also recall the condition $({b_1^j}^+a_2^j)-({b_2^j}^+a_1^j)=2.$
The dimension of the reduced space is $4Nl+2N-2l(l-1)-N-N=4Nl-2l(l-1).$
We fix $W_1,W_2$ such that the matrices
\BEQ
W_1^{-1} \sum_i a_1^i {b_2^i}^+ W_1,\qquad W_2^{-1}
\sum_i a_2^i {b_1^i}^+W_2
\label{eq:matr}
\EEQ
are diagonal and the matrices $W_k$ leave the vector $(1,\ldots,1)$
invariant. Further we use the notation
$A_i^k,B_j^m$ for the new variables. The normalization
 (\ref{eq:matr}) takes the form:
$$ \sum_i A_{il}^1 B_{im}^2 =\delta_{lm} k_l^1,$$
$$ \sum_i A_{il}^2 B_{im}^1 =\delta_{lm} k_l^2.$$
For the eigenvector of the operator $L$ the
following representation is true:
$$c_i=(z-z_1)(c_i^j+O(z-z_1))\exp(-x_i\zeta(z)), \mbox{ if }j \leq l,$$
$$
c_i=(z-z_2)(c_i^j+O(z-z_2))\exp(-x_i\zeta(z)), \mbox{ if }2l\geq j>l,$$
$$
c_i=(c_i^j+O(z))\exp(-x_i\zeta(z)),\mbox{ if }j>2l.$$
The vectors $c^j$ are proportional to the vectors
$B^j$. We chose for normalizing the following condition
\BEQ
\sum_{i=1}^N A_{im} c_i^j =\delta_{jm}, j \leq 2l,\qquad
\sum_{i=1}^N A_{im} c_i^j =0,j>2l.
\label{cond}
\EEQ
It implies the customary normalizing of FBA at the points
$P_j$
$$\Psi_q =(\delta_{jq}+ O(z))\exp(k_j x).$$
Now we define the ``action'' variables as the poles
of the vector $C$ which solves the equation
$(k+L)C=0$ and satisfies the normalizing condition
\BEQ
\sum_{m=1}^{l} \sum_{i=1}^N A_{im} c_i \Phi_1(-x_i,z)+
\sum_{m=l+1}^{2l} \sum_{i=1}^N A_{im} c_i \Phi_2(-x_i,z)=1
\label{nor}
\EEQ
which is completely in agreement with the previous normalization due to
Riemann-Roch theorem. Indeed, the expression on the left-hand side is
a meromorphic function on the spectral curve, it has $g+2l-1$ poles and
it has fixed values $1$ at the points $P_i$, i.e. additional
$2l$ conditions. This function has to be equal to $1$ identically.
Now all the preliminaries are done and we turn to the construction.
The Lax operator and its eigenvector
normalized by the condition (\ref{nor})
are natural coordinates on the phase space and we take
as $\delta L$ and $\delta C$ the corresponding cotangent
vectors.
We only introduce new variables $k^{new},L^{new}$ such that
$k^{new}=\hat k-2\zeta(z)+\zeta(z-z_1)+\zeta(z-z_2)$ which
is a well defined function on the spectral curve, new variables
$\lambda_i^{new}=\lambda_i-\zeta(\eta)(({B_2^i}^+ A_1^i)+({B_1^i}^+ A_2^i))$
and the corresponding
expression for the Lax operator:
$$
L^{new}_{ij}=\delta_{ij}( \lambda_i^{new}
+\zeta(z-z_2)(({B_1^i}^+ A_2^i)-1)-\zeta(z-z_1)(({B_2^i}^+ A_1^i)+1))+
$$
\BEQ
(1-\delta_{ij})(({B_1^i}^+ A_2^i)\Phi_2(x_i-x_j,z)-({B_2^i}^+ A_1^i)
\Phi_1(x_i-x_j,z)).
\label{Lax}
\EEQ
The index $new$ is omitted in the following.
We also introduce the differentials on the
total space
$\delta k$ and $\delta z$.
To relate it with the canonical differential on the
spectral curve we notice that the total space is the
fibration on the specific moduli space with fiber which is the symmetric
power of the spectral curve. We also notice that
the sub-varieties  $z=const$ are transversal
to the fibers. So we could extend
differentials on the fiber to the total space
defining them zero along the sub-varieties $z=const$.
It means that we choose the connection which is zero along
such sub-varieties. For this choice of a connection we have
$\delta z = dz$, where we use the notation $d$ for the
extended differential (for more details see \cite{KP,KP1}).

The main object of our concern is the two-form on the phase space
\BEQ
\omega = \frac 1 2 \sum_{i=1}^N   res_{P_i} (<C^* (\delta L + \delta k )
\wedge  \delta C >) d z,
\label{uni}
\EEQ
where $C^*$ is the eigenvector of $L$ satisfying $<C^* C>=1$.
\begin{pr}
The  two-form $\omega$ is
\BEQ
\omega = 2 \sum_s \delta z ( \gamma_s )
\wedge \delta k( \gamma_s ).
\label{act}
\EEQ
\end{pr}
{\bf Proof}
\quad We first clarify the meaning of this formula. $k$ is a well defined
function on the curve and $z$ is a multi-valued function. After
the choice of a branch of the function $z$,
its values at the points $\gamma_s$ are functionals
on the space of spectral data, i.e. on the phase space of our system.
The only thing to justify is that the corresponding cotangent
vector $\delta z (\gamma_s)$ does not depend on the choice of branch.
Indeed, the difference between such functionals on different
branches depends only on the modular parameter $\tau$ of the
base elliptic curve which is a priori fixed.
The differential $\Omega=<C^*\delta L \wedge \delta C>dz$ is meromorphic
on the spectral curve, the essential singularities reduce.
The sum of the residues at the points $P_i$ is the opposite
of the sum of other residues. $\Omega$ has poles at
the points $\gamma_s$. Using the condition $<C^* C>=1$ we
obtain that the residues at these points are
$$res_{\gamma_s}\Omega=-<C^*\delta L C>\wedge \delta z(\gamma_s)=
-\delta k(\gamma_s) \wedge \delta z(\gamma_s).$$
Other possible poles of $\Omega$ are the branch points of the curve,
at these points $C^*$ has poles due to the normalization,
but the differential $dz$ has
simple zeroes at these points.
$\blacksquare$

This proposition shows that our two-form is symplectic, i.e.
it is closed and non-degenerate. It is non-degenerate because
the number of Darboux coordinates is equal to the dimension
of the phase space.
\begin{pr} The symplectic form $\omega$ admits the
representation
\BEQ
\omega =  \sum_{i=1}^N (\delta \lambda_i \wedge \delta x_i +
\delta  {B_i^q}^1 \wedge \delta A_{iq}^2 -
 \delta  {B_i^q}^2 \wedge \delta A_{iq}^1 ).
\label{AB}
\EEQ
\end{pr}
This is proven by a straightforward calculation. We use
the notation:
$$A_i=(A^2_{i1},\cdots,A^2_{il},-A^1_{i1},\cdots,-A^1_{il});$$
$$B_i=(B^1_{i1},\cdots,B^1_{il},B^2_{i1},\cdots,B^2_{il}).$$
Let us decompose $L,C$ as $L=G \tilde L G^{-1},C=G \tilde C$
where $G=Diag(e^{\zeta(z)x_i}).$ Then
$$\omega=\frac 1 2 res_{z=0}Tr[\delta \tilde L \wedge \delta h+
\hat C^{-1}(\delta \tilde L \delta \hat C +[\delta h,\tilde L]
\wedge \delta \hat C-\delta h \hat C \wedge \delta \hat k)].$$
Here the expression $\hat C$
corresponds to the section of the direct image of the line bundle
on the spectral curve to the base elliptic curve
and
$\delta h=G^{-1}\delta G=diag(-\delta x_i \zeta(z)).$
Using
$$Tr(C^{-1}[\delta h,L]\wedge \delta C)=-Tr(C^{-1}\delta h \wedge
(\delta L \hat C - \hat C \delta \hat k)),$$
we obtain
$$\omega=\frac 1 2 res_{z=0}Tr(2\delta \tilde L \wedge \delta h+
\hat C^{-1}\delta L \wedge \delta C).$$
The first summand gives
$\sum_i \delta \lambda_i \wedge \delta x_i.$
Calculating the second summand we take into account the
normalizing conditions (\ref{cond})
$$\sum_{j=1}^N (A_{j\alpha}\delta c_j^k+\delta A_{j\alpha}c_j^k)=0,
\qquad \sum_{j=1}^N(c_{kj}^*\delta B_{j\alpha}+\delta c_{kj}^*
B_j^{\alpha})=0,$$
which demonstrates the proposition.
$\blacksquare$
\begin{pr} The system of poles dynamics of
the finite gap potentials of the
Davey-Stewartson equation (\ref{El1}),...,(\ref{El6}) is hamiltonian
with respect to the symplectic form $\omega$ and the
Gaudin Hamiltonian
$$
H=\frac 1 2 \sum_i \lambda_i ({B_2^i}^+ A_1^i+{B_1^i}^+ A_2^i)
-\sum_i \zeta(\eta) ({B_2^i}^+ A_1^i+1)({B_1^i}^+ A_2^i-1)
+
$$
\BEQ
\sum_{i \ne j}
({B_1^i}^+ A_2^j) ({B_2^i}^+ A_1^i)
\sphi{x_j-x_i}{+}{\eta}
\label{ham1}
\EEQ
is equal to $(H_1-H_2)/2$, defined in section 1 for the Gaudin elliptic
system.
\end{pr}
{\bf Proof}
\quad We substitute the flow defining the dynamics, i.e. the Lax equation,
into the symplectic form.
Our intention is to show that $i_{\partial_t} \omega (X) = \omega (X,
\partial_t)= dH(X).$ With the chosen
coordinates the vector field is defined by the expressions:
$$\partial_t (L_1-L_2) = [L_1,L_2],\qquad
\partial_t k =0;$$
$$\partial_ t C(t,P) = L_2 C(t,P) +\mu(t,P) C(t,P),$$
where $\mu(t,P)$ is a scalar function.
$$
i_{\partial_t}  \omega = \frac 1 2 \sum_{i=1}^{2l} res_{P_i}
(<C^* (\delta L-\delta k) (L_2 +\mu(t,P)) C> -
<C^*[L_2 ,L_1] \delta C>) dz=
$$
\BEQ
=-\sum_{i=1}^{2l} res_{P_i} \delta k \mu(t,P) dz.
\label{eq:58}
\EEQ
The function $\mu$ has the same expansion at
the marked points as the coefficient at $t$ in
the exponential part of the vector $\Psi$.
It means that at $P_i,i=1...l$ the principal part of $\mu$
is equal to $ \frac 1 2 k$ and at the points
$P_i,i=l+1,...,2l, \mu \approx - \frac 1 2 k$.
$$
i_{\partial_t} \omega =-\frac 1 4 \sum_{i=1}^l res_{P_i} \delta k^2 dz +
\frac 1 4 \sum_{i=l+1}^{2l} res_{P_i} \delta k^2 dz =
$$
$$
= -\frac 1 4 \sum_{i=1}^l res_{P_i} \delta <C^*L^2C> dz+
\frac 1 4 \sum_{i=l+1}^{2l} res_{P_i} \delta <C^*L^2C> dz=
$$
$$
=\frac 1 4 res_{z=z_2} Tr \delta L^2 dz -\frac 1 4
res_{z=z_1} Tr  \delta L^2 dz.
$$
Substituting the expressions for the matrix elements
of the Lax operator (\ref{Lax}) into this formula we find
$$
res_{z=z_2}Tr L^2 dz -res_{z=z_1} Tr L^2 dz =
$$
$$
\frac 1 2 \sum_i (\lambda_i (({B_2^i}^+ A_1^i)+({B_1^i}^+ A_2^i))+
\zeta(\eta)(({B_2^i}^+ A_1^i)^2+({B_1^i}^+ A_2^i)^2))+
$$
$$
\sum_i \zeta(\eta) (({B_2^i}^+ A_1^i)-({B_1^i}^+ A_2^i)+1)+
\sum_{i \ne j} ({B_1^i}^+ A_2^j) ({B_2^i}^+ A_1^i)
\sphi{x_j-x_i}{+}{\eta}.
$$
We have found the Hamiltonian up to a constant. To obtain the
Hamiltonian $H_1-H_2$ of the Gaudin system we have to subtract $N\zeta(\eta)$
and use the condition $({B_1^i}^+ A_2^i)-({B_2^i}^+ A_1^i)=2$.
We only need to note that the system (\ref{El1}),...,(\ref{El6}) is
equivalent to the system of the hamiltonian
dynamics with this Hamiltonian,
and the equivalence is given by the formulas:
$$
a_k^i \mapsto \exp(-\int(\lambda_1^i+\lambda_2^i)/2)a_k^i;
$$
$$
b_k^i \mapsto \exp(\int(\lambda_1^i+\lambda_2^i)/2)b_k^i.
$$
$\blacksquare$

The ``action-angle'' variables can be constructed in the traditional way
as in \cite{K2}. We introduce the differentials $d\Omega_j,j=1,\ldots,2l-1$
of the third kind with residues $1$ and $-1$ at the singular points
$P_j$ and $P_{2l}$ respectively. On the fundamental domain of our curve
we can define the functions $A_k(Q),k=g+2l-1$ such that
$$A_k(Q)=\int^Q d\omega_i,i=1,\ldots,g,\qquad
A_{g+i}(Q)=\int^Q d\Omega_i,i=1,\ldots,2l-1.$$
Now we introduce the quantities
$$
\phi_i=\sum_{s=1}^{g+2l-1}A_i(\gamma_s),
\qquad I_k=\oint_{a_i}kdz,\qquad I_{g+i}=res_{P_i}kdz.
$$
The symplectic form is
$$\omega=\sum_{i=1}^{g+2l-1}\delta \phi_i \wedge \delta I_i.$$
Notice that the quantities $I_i$ are integrals of motion. So,
the obtained coordinates are the ``action-angle'' variables for our system.
For the $sl(2)$-case the separated variables was also
realized in \cite{EFR}.


\section{Degenerations}
The construction of the Hitchin system on degenerate curves
is well suited for the analysis of several degenerations of dynamical
systems. For example the rational and trigonometric degeneration
can be obtained by considering the elliptic curve with
cuspidal and nodal singularities respectively instead of the
regular elliptic curve. The normalization in both cases is
the rational curve and the moduli of holomorphic bundles
are parameterized only by the elements gluing the fibers over
the singular points.
For the Gaudin elliptic system
such degenerations exist. The
rational one is represented by the hamiltonian
\BEQ
H_{rat}=\frac 1 2 \sum_i \lambda_i ({B_2^i}^+ A_1^i+{B_1^i}^+ A_2^i)
+ \sum_{i \ne j}
({B_1^i}^+ A_2^j) ({B_2^i}^+ A_1^i)
\left( \frac 1 {x_i-x_j} - \frac 1 \eta  \right)
\label{ratham}
\EEQ
with the canonical symplectic form $\omega$ in the representation
(\ref{AB}). This system involves a rational curve with two
simple points and one double point with cuspidal singularity.
It also could be solved by the algebraic-geometric
direct and inverse problem as it was done in \cite{Tal}, \cite{Tal1}.
The expressions for the $\Psi$ function could be found
in \cite{Tal1} (2.53).
The ``action-angle'' variables for it were also constructed
in \cite{Br}. The author deals with another Hamiltonian
but the separated variables for these systems coincide.
There is another type of degenerations of our original system.
It is the limit $z_1\to z_2,\eta \to 0.$ This case is
successfully taken care of by the direct and inverse problem as well.
The FBA function can be found in \cite{Tal1} (2.60).
The solution
provides the double periodic potentials for the Davey-Stewartson
equation. The system can be specified by the Hamiltonian
\BEQ
H_{ell}=\frac 1 2 \sum_i \lambda_i ({B_2^i}^+ A_1^i+{B_1^i}^+ A_2^i)
+ \sum_{i \ne j}
({B_1^i}^+ A_2^j) ({B_2^i}^+ A_1^i)
\zeta(x_i-x_j)
\label{ratell}
\EEQ
and the standard symplectic form (\ref{AB}).
A priori this system is not integrable.
Proceeding with the analogous construction
for the ``action-angle'' variables, we obtain that their number is
$2l(2N-2l+1)$ which is less than the dimension of the phase space.
The same degeneracy can be observed by counting the
number of Hitchin Hamiltonians. The
Hamiltonians of the original system can be represented as
$H_{s}^{k,l}=res_{z=z_s}((z-z_s)^{k-1}Tr \Phi^l),s=1,2.$ When the
two fixed points
coincide, there is no more residues of the expressions
$Tr \Phi^k$, and the only retrievable quantities are
$H_1^{k,l}+H_2^{k,l}.$ This case should be investigated with a more
delicate geometrical analysis of the jet structure of the
holomorphic bundle.

\section*{Conclusion}
The main result of this paper is the explicit solution
of the Gaudin elliptic system with spin and the uncovered connection
between this system and the quasi periodic solutions of
the Davey-Stewartson equation. For the methodological
aspects we noticed the crucial correspondence between
the Hitchin and finite gap descriptions of the dynamical systems.
Our subsequent concern is to give a more general illustration
of the equivalence of these two settings.
The important technical query in this direction is
the geometry of the moduli space of semi-stable bundles when
we vary the base algebraic curve. It could clarify the
Hitchin description of such a system as an elliptic degenerate
Gaudin model. Such an attempt was undertaken and the case of the
elliptic curve was analyzed in \cite{OL}. The construction
crucially used the group structure on the elliptic curve
and could not be translated directly to the general situation.
In order to realize the geometrically established limit procedure
a kind of canonical connection on the bundle over the
moduli space of algebraic curves with moduli space of
holomorphic bundles as the fiber is required. And the
role of the Painlev\'e connection in this
context is of immediate concern to us.

\vskip 1.0cm
{\em Acknowledgements}

I would like to thank I. Krichever for introducing me to the
beauty of integrable systems and the special spectral method and
for his proposal to investigate the special spectral problem for
the Davey-Stewartson equation. I would like to thank M.
Olshanetsky and A. Levin for fruitful discussions on the nature of
the Hitchin system. This work was partially supported by the RFBR
01-01-00546 grant.


\newpage
\begin{thebibliography}{50}
\bibitem{NN}
N. Nekrasov, {\em Holomorphic bundles and many-body systems.}
PUPT-1534, Comm. Math. Phys.,{\bf 180} (1996) 587-604;
hep-th/9503157.

\bibitem{ER1}
B. Enriquez, V. Rubtsov, {\em Hitchin systems, higher Gaudin
operators and $r$-matrices.} Math. Res. Lett. {\bf 3} (1996)
343-357; alg-geom/9503010.

\bibitem{K1}
I. Krichever, {\em Nonlinear equations and elliptic curves.} Modern
problems in mathematics, Itogi nauki i tekhniki, VINITI AN USSR
{\bf23} (1983).

\bibitem{GK}
A.Gorsky, I.Krichever, A.Marshakov, A.Mironov, A.Morozov
{\em Integrability and Seiberg-Witten Exact Solution.}
hep-th/9505035

\bibitem{DW}
R. Donagi, E. Witten, {\em Supersymmetric Yang-Mills theory and integrable systems,}
Nucl. Phys. {\bf B} 460 (1996) 299-334.

\bibitem{KBBT}
I. Krichever, O. Babelon, E. Billey, M. Talon, {\em Spin generalization
of the Calogero-Moser system and the Matrix KP equation.} Preprint LPTHE 94/42.
Amer. Math. Transl. {\bf 170} (1995) N{\bf 2}, 83-119.

\bibitem{KZ}
I. Krichever, A. Zabrodin, {\em Spin generalization of the model Ruijsenaars-
Schneider, non-abelian 2-dimensional Toda chain and the Sklyanin
algebras representations.} Uspehi mat. nauk. 1995 V.{\bf 50} N.{\bf 6} 3-56.

\bibitem{Hit}
N. Hitchin, {\em Stable bundles and integrable systems.} Duke Math. Journal
1987 V {\bf 54} N{\bf 1} 91-114.

\bibitem{GP}
Bert van Geemen, Emma Previato,
{\em On the Hitchin System.}
Duke Math. J. {\bf85} (1996) 659-684.

\bibitem{GAW}
Krzysztof Gawedzki, Pascal Tran-Ngoc-Bich,
{\em Hitchin Systems at Low Genera.}
J.Math.Phys. {\bf41} (2000) 4695-4712.



\bibitem{Mal}
T. Malanyuk, {\em Finite-gap solutions of Davey-Stewartson equation.}
Usp. Mat. Nauk. V.{\bf 46} N.{\bf 5} 171-172.

\bibitem{Tal}
D. Talalaev, {\em Finite-gap potentials of the Dirac equation and the related
dynamical systems.} Journal of Math. Phys. V{\bf6} 1999 N{\bf4} 471-492.

\bibitem{Tal1}
D. Talalaev,  {\em $R$-matrix formalism and the finite gap
integrability.} PhD Thesis, {\small \sf
http://wwwth.itep.ru/mathphys/people/talalaev/phd.htm}
in Russian.

\bibitem{KP}
I. M. Krichever, D. H.Phong, {\em On the integrable geometry of soliton
equations and $N=2$ supersymmetric gauge theories.} J. Differential Geometry
{\bf45} (1997) 349-389.

\bibitem{KP1}
I. M. Krichever, D. H. Phong, {\em Symplectic forms in the theory of solitons.}
preprint hep-th/9708170.

\bibitem{K2}
I. Krichever, {\em Elliptic solutions to difference non-linear equations
and nested Bethe Ansatz equation.} solv-int/9804016.

\bibitem{EFR}
B. Enriquez, B. Feigin, V. Rubtsov,
{\em Separation of variables for Gaudin-Calogero systems.}
q-alg/9605030.

\bibitem{Br}
T. Brzezi\'nski, {\em Dynamical $r$-matrices and separation of variables:
the generalized Calogero-Moser model.} hep-th/9401049.

\bibitem{OL}
M. Olshanetsky,
{\em Generalized Hitchin systems and Knizhnik-Zamolodchikov-Bernard
equation on elliptic curves.} Lett. Math. Phys. {\bf42} (1997) 59-71.

\end{thebibliography}
\end{document}